\documentclass[fleqn,usenatbib]{mnras}

\usepackage[T1]{fontenc}

\DeclareRobustCommand{\VAN}[3]{#2}
\let\VANthebibliography\thebibliography
\def\thebibliography{\DeclareRobustCommand{\VAN}[3]{##3}\VANthebibliography}

\usepackage{graphicx}	
\usepackage{amsmath}	
\usepackage{amssymb}	
\usepackage{fontawesome}
\usepackage[dvipsnames]{xcolor}

\usepackage{newtxtext,newtxmath}

\newcommand{\au}{\,\textsc{au}}

\title[Hierarchical triple-star stability]{Algebraic and machine learning approach to hierarchical triple-star stability}

\author[Vynatheya et al.]{
Pavan Vynatheya$^{1}$, Adrian S. Hamers$^{1}$, Rosemary A. Mardling$^{2}$ and Earl P. Bellinger$^{1,3}$
\\
$^{1}$Max-Planck-Institut für Astrophysik, Karl-Schwarzschild-Straße 1, 85748 Garching bei München, Germany \\
$^{2}$School of Physics and Astronomy, Monash University, Clayton Victoria 3800, Australia \\
$^{3}$Stellar Astrophysics Centre, Department of Physics and Astronomy, Aarhus University, Ny Munkegade 120, Aarhus, Denmark
}

\date{Accepted XXX. Received YYY; in original form ZZZ}

\pubyear{2022}

\begin{document}
\label{firstpage}
\pagerange{\pageref{firstpage}--\pageref{lastpage}}
\maketitle

\begin{abstract}
We present two approaches to determine the dynamical stability of a hierarchical triple-star system. The first is an improvement on the Mardling-Aarseth stability formula from 2001, where we introduce a dependence on inner orbital eccentricity and improve the dependence on mutual orbital inclination. The second involves a machine learning approach, where we use a multilayer perceptron (MLP) to classify triple-star systems as `stable' and `unstable'. To achieve this, we generate a large training data set of $10^6$ hierarchical triples using the $N$-body code MSTAR. Both our approaches perform better than previous stability criteria, with the MLP model performing the best. The improved stability formula and the machine learning model have overall classification accuracies of $93\%$ and $95\%$ respectively. Our MLP model, which accurately predicts the stability of any hierarchical triple-star system within the parameter ranges studied with almost no computation required, is publicly available on Github in the form of an easy-to-use Python script.
\end{abstract}


\section{Introduction} \label{sec:intro}

The three-body problem has been of interest to physicists ever since classical mechanics was formulated. Unlike the two-body problem, the equations governing the motion of three bodies do not admit general closed-form solutions; the fact that solutions exist that exhibit extreme sensitivity to initial conditions is evidence that this must be true. There are a few specific cases, including the test particle limit or the `restricted' three-body problem, for which closed-form analytic solutions exist, but the general problem can be chaotic and can only be solved numerically. In the past few decades, the advent of computing has vastly improved our knowledge of $N$-body dynamics.

In this paper, we shall focus on hierarchical triple-star systems. Such a hierarchical configuration can be regarded as two `nested' binaries, with an inner binary of two stars being orbited by a third companion. Unlike isolated binaries, the orbits of a triple can change over periods much greater than orbital timescales, and significantly, depending on the system. This secular evolution in triples leads to, in the lowest-order approximation, von Zeipel-Lidov-Kozai (LK or ZLK) oscillations (\citealp{1910AN....183..345V,1962P&SS....9..719L,1962AJ.....67..591K}; see \citealp{2016ARA&A..54..441N} for a review), which are periodic changes seen in the inner orbital eccentricity and the mutual orbital inclination. Drastic changes in eccentricity can lead to stellar collisions if the orbital periapsis is on the order of stellar size and the LK oscillations are not quenched by general relativistic or tidal effects. Additionally, a large eccentricity can mark the onset of dynamical instability and the subsequent escape of one of the stars.

The physical origin of chaotic behaviour in hierarchical triples involves resonant interactions between the inner orbit and harmonics of the outer orbit \citep{2008LNP...760...59M,2013MNRAS.435.2187M}. In particular, significant energy exchange between the orbits must take place for one of the stars to escape, and this can only occur if there is enough power in the harmonic whose frequency is close to the inner orbital frequency. This in turn requires sufficiently high outer eccentricity. As a result, for a particular integer $N\simeq P_{\rm out}/P_{\rm in}$, it is the $N:1$ `resonance' and its neighbours which determine stability in triple star systems, with the \textit{resonance overlap stability criterion} from \cite{1979PhR....52..263C} providing a powerful heuristic for assessing this \citep{2008LNP...760...59M}. Resonance is at the heart of the simpler \cite{2001MNRAS.321..398M} stability criterion (see Equation~\ref{eqn:MA_form} below), which is based on the proposal that the ratio of outer to inner semimajor axes is proportional to some power of the ratio of the time of outer periastron passage to the inner orbital period.

Investigating the stability of hierarchical triples is important because of their abundance in the universe. \cite{2017ApJS..230...15M} found that $\gtrsim 50\%$ of massive O- and B-stars reside in multiple-star systems like triples and quadruples, compared to $\lesssim 10\%$ of solar-mass stars. Hence, the study of massive stars, which includes high-energy phenomena like supernovae and gravitational waves, is incomplete without first understanding the dynamical evolution of triples. However, direct $N$-body integration is computationally expensive and not always desirable. This makes it necessary to come up with techniques to predict the long-term stability of hierarchical configurations.

Besides being compelling from a pure dynamics perspective, the question of hierarchical triple-star stability is of interest in triple population synthesis studies (e.g., \citealp{2016ApJ...816...65A, 2019MNRAS.486.4443F}). In this context, dynamical stability criteria are an important step in the initial sampling of a population of triple-star systems. A poor classification can impact the statistics of the problem.

In this work, we compare our results to the stability criteria presented by \cite{1995ApJ...455..640E} and \cite{2001MNRAS.321..398M}. \cite{1995ApJ...455..640E} proposed an empirical fit for the stability of hierarchical triple-star systems. They assumed a system to be stable if it remains in the same hierarchical configuration after 100 orbits of the `outer' binary. A more widely used stability criterion was provided by \cite{2001MNRAS.321..398M} (see also \citealp{Mardling1999}). They drew parallels between dynamical instability in triples and tidal evolution and presented a semi-analytical formula to distinguish between stable and unstable systems.

The structure of this paper is as follows. Section~\ref{sec:nbody} briefly describes the $N$-body code we used and the assumptions we make about stability; Section~\ref{sec:data} details the initial conditions and our parameter space; Sections \ref{sec:form} and \ref{sec:mlp} form the crucial components of this paper, describing our updated stability criterion and our machine learning classifier respectively; Section~\ref{sec:result} summarises our results; Section~\ref{sec:discuss} is the discussion; and Section~\ref{sec:conclude} finally concludes. Appendix \ref{sec:model} provides a brief tutorial to use our machine learning model.

\section{\textit{N}-body code and stability} \label{sec:nbody}
For our study of hierarchical triple-star systems, we use the $N$-body code MSTAR (see \citealp{2020MNRAS.492.4131R} for details), which performs highly accurate integration for a wide range of mass ratios. For simplicity, we use the code with the post-Newtonian (PN) terms disabled. The PN terms have little effect when distances are in the order of $\sim 1 \au$ and for stellar-mass scales but can be significant during close encounters and compact object mergers. The reason for ignoring PN terms is to make our problem \textit{scale-free} in mass, distance and time. Thus, we do not need to be concerned about the actual values of masses and distances and instead, concentrate on ratios of quantities.

With the scale-free assumption, hierarchical triple systems have the following relevant initial parameters, which dictate the evolution of the system:

\begin{itemize}
    \item Inner mass ratio $q_{\mathrm{in}} = m_2/m_1 \le 1$ ($m_2 \le m_1$), where $m_1$ and $m_2$ are the inner binary masses of the hierarchical triple.
    \item Outer mass ratio $q_{\mathrm{out}} = m_3/(m_1+m_2)$, where $m_3$ is the outer mass of the heirarchical triple.
    \item Semimajor axis ratio $\alpha = a_{\mathrm{in}}/a_{\mathrm{out}} < 1$, where $a_{\mathrm{in}}$ and $a_{\mathrm{out}}$ are the semimajor axes of the inner and outer orbits of the hierarchical triple respectively.
    \item Inner orbit eccentricity $0 \le e_{\mathrm{in}} < 1$.
    \item Outer orbit eccentricity $0 \le e_{\mathrm{out}} < 1$.
    \item Mutual inclination $i_{\mathrm{mut}}$ between the two orbits of the hierarchical triple.
\end{itemize}

Here, the mutual inclination $\cos{(i_{\mathrm{mut}})} = \cos{(i_{\mathrm{in}})} \cos{(i_{\mathrm{out}})} + \sin{(i_{\mathrm{in}})} \sin{(i_{\mathrm{out}})} \cos{(\Omega_{\mathrm{in}} - \Omega_{\mathrm{out}})}$ takes into account two of the three orbit-orientation (Euler) angles, with the inclination $i$ and the longitude of ascending node $\Omega$ being defined with respect to a chosen reference direction (the subscripts refer to the inner and outer orbits). In fact, as demonstrated in panels (e) and (f) of Figure 3.16 of \cite{2008LNP...760...59M}, the stability boundary is also sensitive to the third Euler angle, the argument of periapsis $\omega$, since it is this which determines the position of a system relative to the centre of the relevant $N$ : 1 resonance, and as a consequence, the system's stability. We do not attempt to capture the resulting step-like structure of the stability boundary, instead adopting an expression which effectively smooths this behaviour. Similarly, any dependence on the true anomaly $\theta$ is not captured in this work.

\begin{figure}
	\includegraphics[width=1.0\columnwidth]{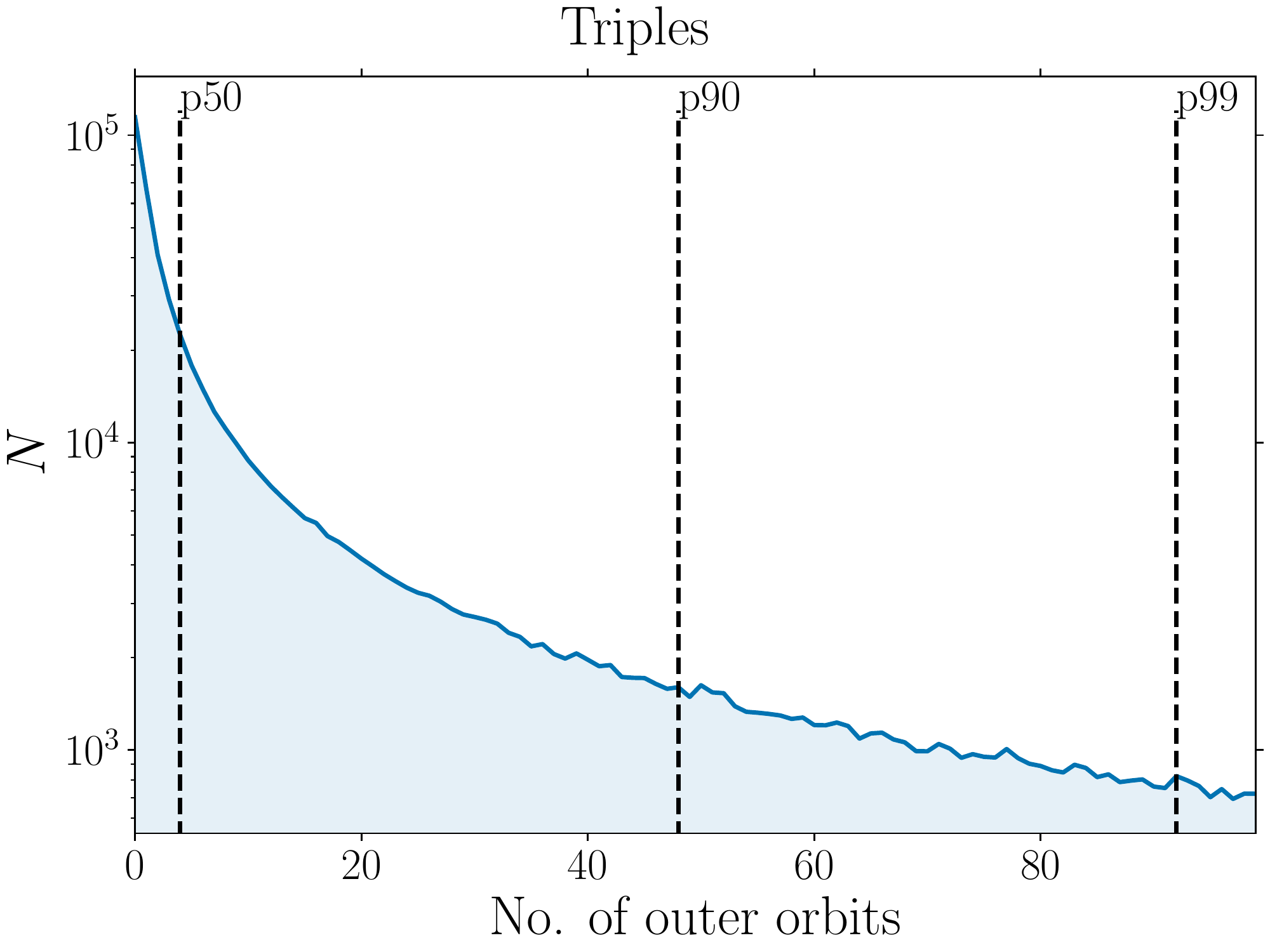}
    \caption{Plot of the number of outer orbits within which unstable triple-star systems become unbound. The dashed lines represent the 50th, the 90th and the 99th percentile values, which shows that the fraction of unstable systems that become unbound within 100 outer orbits decreases rapidly.}
    \label{fig:boundorbits}
\end{figure}

We now come to the defining criterion for the stability of a triple. Newtonian mechanics predicts that a bound two-body system remains in a closed elliptical orbit indefinitely. On the other hand, the evolution of a three-body system can be, in general, chaotic and unpredictable. In many cases, minute changes in the initial parameters can manifest as substantial differences in evolution in secular timescales. Thus, we need to decide about how to quantify stability. We deem a triple system stable if it remains bound for 100 outer orbits and if the semimajor axes of both inner and outer orbits do not change by more than 10\% of the initial value. The rationale for choosing 100 orbits is shown in Figure~\ref{fig:boundorbits}. The most unstable systems become unbound well before 100 outer orbits. There exist a small fraction of unstable systems which remain bound after 100 outer orbits, and the second criterion applies to these systems. The 10\% threshold for semimajor axis change ensures that triple systems on the verge of becoming unstable are not erroneously classified as stable.

\section{Data set and initial conditions} \label{sec:data}
We generate initial conditions for hierarchical triple systems such that the parameter space is filled more or less uniformly. This is necessary to ensure good classification.

Moreover, to analyse the parameter space more thoroughly, we look at different parameter space slices, where one or more initial parameters are kept constant while others are varied. We perform three kinds of parameter space slices:
\begin{enumerate}
    \item Varied semimajor axes and masses, constant eccentricities and mutual inclinations (see Table~\ref{tab:a-m}).
    \item Varied semimajor axes and eccentricities, constant masses and mutual inclinations (see Table~\ref{tab:a-e}).
    \item Varied semimajor axes and mutual inclinations, constant masses and eccentricities (see Table~\ref{tab:a-i}).
\end{enumerate}

\begin{table}
	\centering
	\begin{tabular}{llll}
        \hline
        Slice & $e_{\mathrm{in}}$ & $e_{\mathrm{out}}$ & $i_{\mathrm{mut}}$ \\
        \hline
        Fiducial & 0.0 & 0.0 & 0.0 \\
        $i_{\mathrm{mut}} = \pi/2$ & 0.0 & 0.0 & $\pi/2$ \\
        $i_{\mathrm{mut}} = \pi$ & 0.0 & 0.0 & $\pi$ \\
        $e_{\mathrm{out}} = 0.3$ & 0.0 & 0.3 & 0 \\
        $e_{\mathrm{out}} = 0.3$ & 0.0 & 0.6 & 0 \\
        $e_{\mathrm{in}} = 0.3$ & 0.3 & 0.0 & 0 \\
        $e_{\mathrm{in}} = 0.6$ & 0.6 & 0.0 & 0 \\
        \hline
    \end{tabular}
	\caption{Parameter space slices where $\alpha$, $q_{\mathrm{in}}$ and $q_{\mathrm{out}}$ are varied, and other parameters are kept constant. The columns in the table show the constant values of these parameters in different slices.}
	\label{tab:a-m}
\end{table}

\begin{table}
	\centering
	\begin{tabular}{llll}
        \hline
        Slice & $q_{\mathrm{in}}$ & $q_{\mathrm{out}}$ & $i_{\mathrm{mut}}$ \\
        \hline
        Fiducial & 1.0 & 0.5 & 0.0 \\
        $i_{\mathrm{mut}} = \pi/2$ & 1.0 & 0.5 & $\pi/2$ \\
        $i_{\mathrm{mut}} = \pi$ & 1.0 & 0.5 & $\pi$ \\
        $q = 0.1, 0.818$ & 0.1 & 0.818 & 0.0 \\
        $q = 1, 9$ & 1.0 & 9.0 & 0.0 \\
        $q = 0.8, 0.111$ & 0.8 & 0.111 & 0.0 \\
        \hline
    \end{tabular}
	\caption{Parameter space slices where $\alpha$, $e_{\mathrm{in}}$ and $e_{\mathrm{out}}$ are varied, and other parameters are kept constant. The columns in the table show the constant values of these parameters in different slices.}
	\label{tab:a-e}
\end{table}

\begin{table}
	\centering
	\begin{tabular}{lllll}
        \hline
        Slice & $q_{\mathrm{in}}$ & $q_{\mathrm{out}}$ & $e_{\mathrm{in}}$ & $e_{\mathrm{out}}$ \\
        \hline
        Fiducial & 1.0 & 0.5 & 0.0 & 0.0 \\
        $e_{\mathrm{out}} = 0.3$ & 1.0 & 0.5 & 0.0 & 0.3 \\
        $e_{\mathrm{out}} = 0.3$ & 1.0 & 0.5 & 0.0 & 0.6 \\
        $e_{\mathrm{in}} = 0.3$ & 1.0 & 0.5 & 0.3 & 0.0 \\
        $e_{\mathrm{in}} = 0.6$ & 1.0 & 0.5 & 0.6 & 0.0 \\
        $q = 0.1, 0.818$ & 0.1 & 0.818 & 0.0 & 0.0 \\
        $q = 1, 9$ & 1.0 & 9.0 & 0.0 & 0.0 \\
        $q = 0.8, 0.111$ & 0.8 & 0.111 & 0.0 & 0.0 \\
        \hline
    \end{tabular}
	\caption{Parameter space slices where $\alpha$ and $i_{\mathrm{mut}}$ are varied, and other parameters are kept constant. The columns in the table show the constant values of these parameters in different slices.}
	\label{tab:a-i}
\end{table}

We also limit our parameter ranges to remain in the triple-star system domain:
\begin{itemize}
    \item $10^{-2} \le q_{\mathrm{in}} \le 1$, $10^{-2} \le q_{\mathrm{out}} \le 10^{2}$
    \item $10^{-4} < \alpha < 1$
    \item $0 \le e_{\mathrm{in}} < 1$, $0 \le e_{\mathrm{out}} < 1$
    \item $0 \le i_{\mathrm{mut}} \le \pi$.
\end{itemize}

\section{Updated formula} \label{sec:form}
As mentioned previously, we attempt to improve on the previously existing and widely used stability criterion given by \cite{2001MNRAS.321..398M}.  Before detailing our updated stability criterion, we mention the two stability criteria with which we compare our classification performances.

\begin{itemize}
    \item Defining  $Y = [a_{\mathrm{out}} (1-e_{\mathrm{out}})] / [a_{\mathrm{in}} (1+e_{\mathrm{in}})]$ \cite{1995ApJ...455..640E} (henceforth EK95) derived the following fitting formula for stability:
    \begin{equation}
        Y_{\mathrm{crit}} = 1 + \frac{3.7}{q_{\mathrm{out}}^{-1/3}} - \frac{2.2}{1+q_{\mathrm{out}}^{-1/3}} + \frac{1.4}{q_{\mathrm{in}}^{-1/3}} \frac{q_{\mathrm{out}}^{-1/3}-1}{q_{\mathrm{out}}^{-1/3}+1}.
        \label{eqn:EK_form}
    \end{equation}

    A hierarchical triple-star system is deemed `stable' if $Y > Y_{\mathrm{crit}}$ and `unstable' if $Y < Y_{\mathrm{crit}}$.

    \item The other criterion we will compare against is the semi-analytical, and more accurate, formula by \cite{2001MNRAS.321..398M} (henceforth MA01):
    \begin{equation}
        \frac{R_{\mathrm{p,crit}}}{a_{\mathrm{in}}} = 2.8 \left[ (1+q_{\mathrm{out}}) \frac{1+e_{\mathrm{out}}}{(1-e_{\mathrm{out}})^{1/2}} \right]^{2/5} \left( 1 - \frac{0.3 i_{\mathrm{mut}}}{\pi} \right)
        \label{eqn:MA_form}
    \end{equation}
    
    Here, $R_{\mathrm{p}} = a_{\mathrm{out}} (1-e_{\mathrm{out}})$ is the outer periastron distance and a triple system is deemed `stable' if $R_{\mathrm{p}} > R_{\mathrm{p,crit}}$ and `unstable' otherwise.
\end{itemize}

Finally, we present our stability criterion. After taking into account the dependencies on the initial parameters, we defined a new parameter $\widetilde{e}_{\mathrm{in}}$ as follows:
\begin{equation}
\begin{split}
    e_{\mathrm{in,max}} &= \sqrt{1 - \frac{5}{3} \cos^2{i_{\mathrm{mut}}}}; \\
    e_{\mathrm{in,avg}} &= 0.5 e_{\mathrm{in,max}}^2; \\
    \widetilde{e}_{\mathrm{in}} &= \max(e_{\mathrm{in}}, e_{\mathrm{in,avg}}).
\end{split}
\label{eqn:ein_tilde}
\end{equation}
The first equation is the quadrupole-order approximation for the maximum value of inner eccentricity due to LK oscillations. The second equation describes the \textit{average} eccentricity during an orbit in the sense that the \textit{time-averaged} separation for a single orbit is $\langle r_{\mathrm{in}} \rangle = a_{\mathrm{in}} (1 + e_{\mathrm{in,avg}})$ (see \citealp{doi:10.1080/0025570X.1977.11976639}).

Our updated stability criterion is as follows:
\begin{equation}
\begin{split}
    \widetilde{Y}_{\mathrm{crit}} = 2.4 \left[ \frac{(1+q_{\mathrm{out}})}{(1+\widetilde{e}_{\mathrm{in}}) (1-e_{\mathrm{out}})^{1/2}} \right]^{2/5} \\ \times \left[ \left( \frac{1 - 0.2 \widetilde{e}_{\mathrm{in}} + e_{\mathrm{out}}}{8} \right) (\cos{i_{\mathrm{mut}}} - 1) + 1 \right].
\end{split}
\label{eqn:new_form}
\end{equation}
Here, $\widetilde{Y}$ is the same as $Y$ with $e_{\mathrm{in}}$ replaced by $\widetilde{e}_{\mathrm{in}}$. The second term in the equation accounts for some of the complicated dependence on mutual inclination, whereas the first term is very similar to the MA01 formula with an additional $\widetilde{e}_{\mathrm{in}}$ dependence.

The reasoning behind the choices for the dependencies in Equation~\ref{eqn:new_form} is detailed in the following sub-sections.

\subsection{Eccentricity dependence}
From our parameter space slices of eccentricities, we found that an additional dependence on $e_{\mathrm{in}}$ can better describe the classification boundary between stable and unstable systems. This is unlike the MA01 formula, which has no dependence on $e_{\mathrm{in}}$.

\begin{figure*}
	\includegraphics[width=2.0\columnwidth]{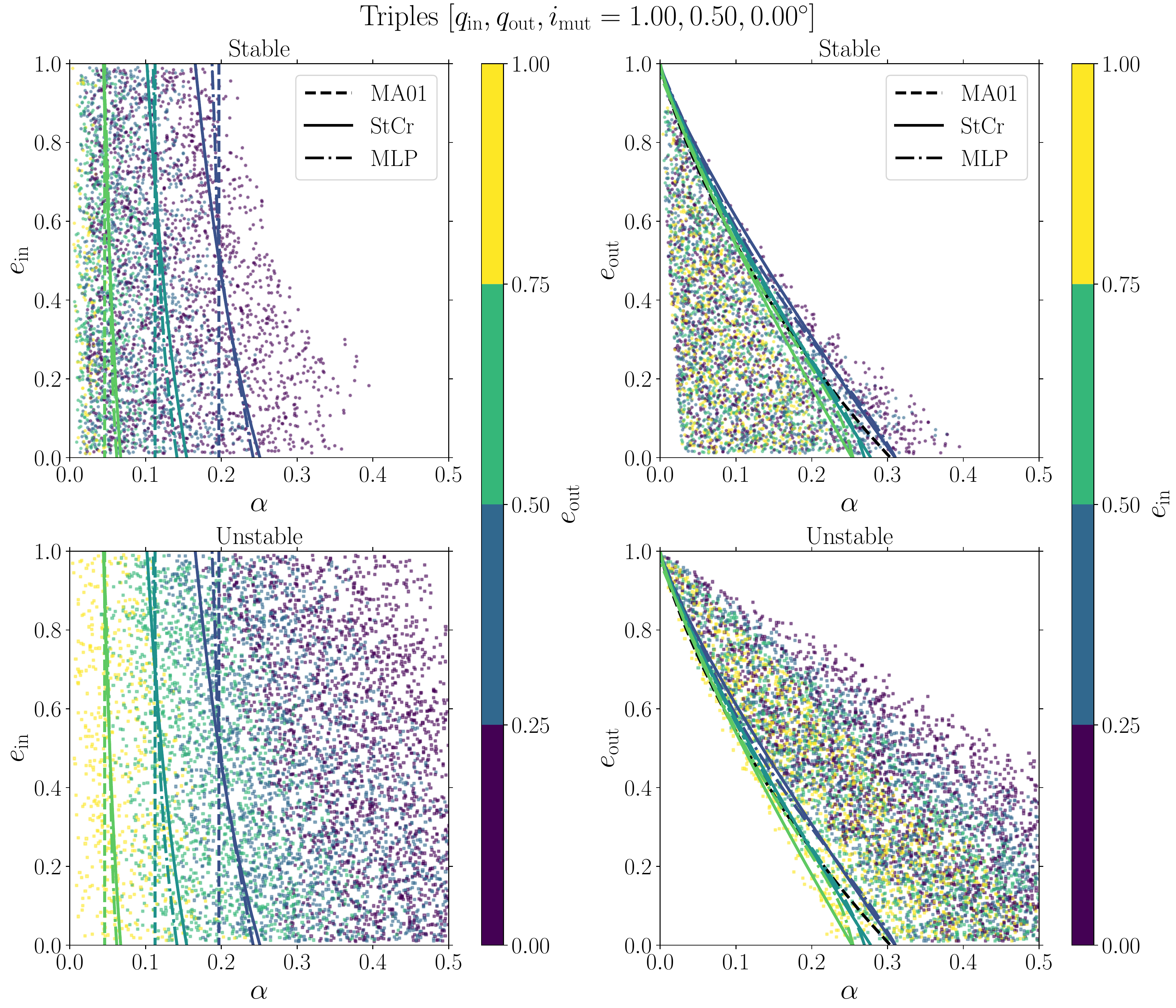}
    \caption{Plot showing a parameter space slice of varying $\alpha$, $e_{\mathrm{in}}$ and $e_{\mathrm{out}}$. The constant values of the other parameters are mentioned at the top. The left panel plots $e_{\mathrm{in}}$ vs $\alpha$, while the right panel plots $e_{\mathrm{out}}$ vs $\alpha$. The missing varied parameter is the colour axis in all panels. The top and bottom panels show the systems which remain stable and become unstable respectively from direct $N$-body simulations. The lines represent contours of the classification boundaries, with the legend labels `MA01', `StCr' and `MLP' referring to the MA01 criterion, our updated stability criterion (Equation~\ref{eqn:new_form}) and our MLP model respectively. The colours of the contour lines coincide with the transition values on the colour axes. The black lines are independent of the colour axis.}
    \label{fig:ae-1}
\end{figure*}

\begin{figure*}
	\includegraphics[width=2.0\columnwidth]{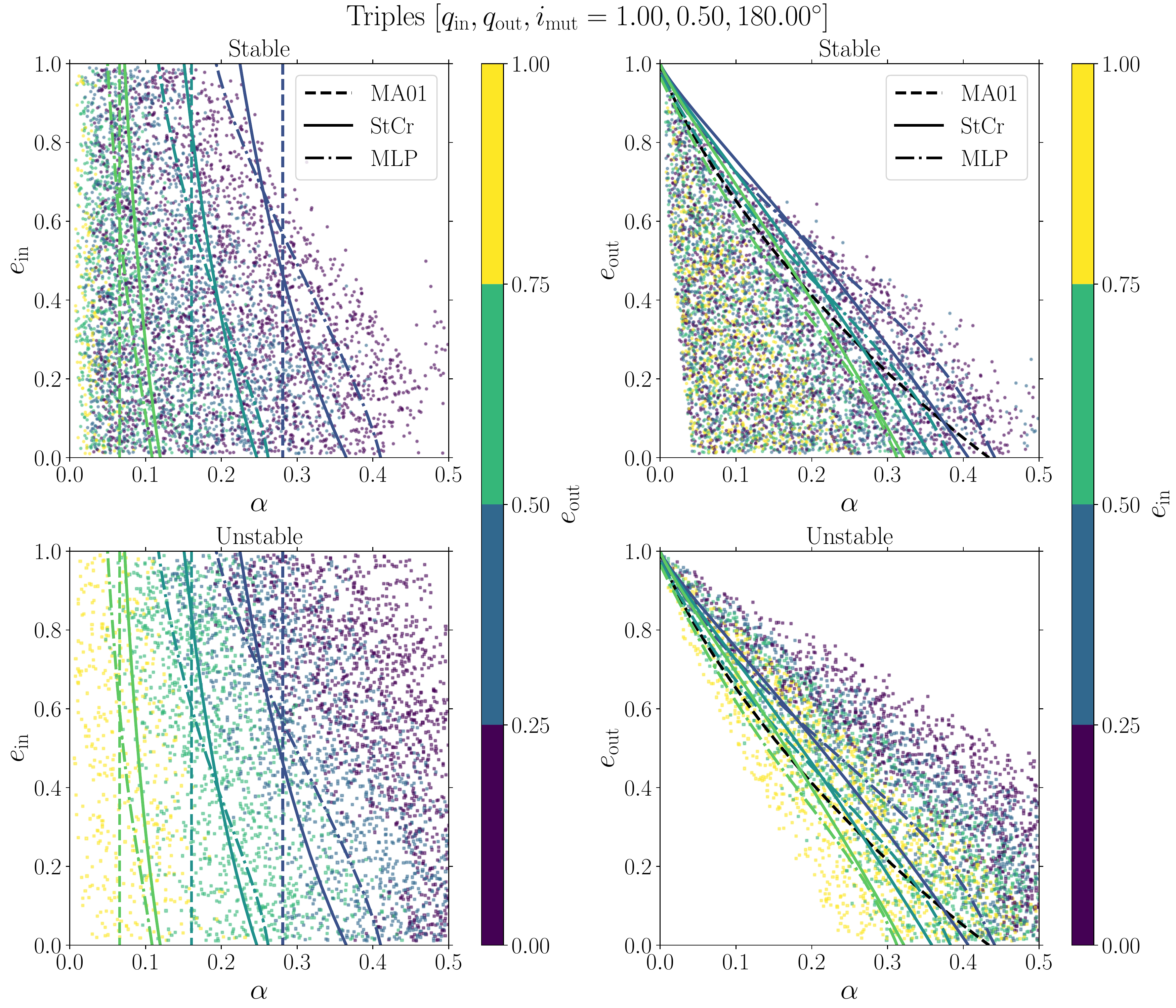}
    \caption{Plot similar to Figure~\ref{fig:ae-1}, but showing a different parameter space slice. The changed constant values are mentioned at the top. }
    \label{fig:ae-3}
\end{figure*}

Figures \ref{fig:ae-1} and \ref{fig:ae-3} illustrate the dependence of stability on both $e_{\mathrm{in}}$ and $e_{\mathrm{out}}$. The dependence of $Y_{\mathrm{crit}}$ on $e_{\mathrm{out}}$ is the same as the MA01 factor $(1 - e_{\mathrm{out}})^{-2/5}$, while the extra dependence on $e_{\mathrm{in}}$ is taken into consideration in Equation~\ref{eqn:new_form} through the factor $(1 + e_{\mathrm{in}})^{-2/5}$. It should also be noted that Equation~\ref{eqn:new_form} works better for prograde orbits (Figure~\ref{fig:ae-1}) than for retrograde orbits (Figure~\ref{fig:ae-3}). Retrograde systems tend to have a stronger dependence on $e_{\mathrm{in}}$ for stability than prograde systems. Nonetheless, it is a notable improvement on previously existing stability criteria.

\subsection{Mutual inclination dependence}
The parameter space slices of mutual inclinations showed that the inclination dependence on stability is not linear. The ad-hoc inclination factor in the MA01 formula, $(1-0.3 i_{\mathrm{mut}}/\pi)$, is a good linear approximation but leaves room for improvement. In particular, systems with mutually highly inclined orbits (around $i_{\mathrm{mut}} = \pi/2$) tend to be less stable than those with both less-inclined prograde and retrograde orbits. Moreover, retrograde orbits tend to be more stable than prograde orbits, which is also captured by the MA01 formula.

To account for the bowl-shaped depression in the stability classification boundary, we employed a trick used by \cite{2017MNRAS.466..276G}. We replaced $e_{\mathrm{in}}$ by $\widetilde{e}_{\mathrm{in}}$, defined in Equation~\ref{eqn:ein_tilde}.

The motivation for doing this substitution is that the inner eccentricity, unlike the outer eccentricity, is not constant in the quadrupole-order approximation of LK oscillations. The $e_{\mathrm{in}}$ value changes the most for highly inclined orbits, and hence, replacing the initial value of $e_{\mathrm{in}}$ with a more average value can help to better characterise such systems.

\begin{figure}
	\includegraphics[width=1.0\columnwidth]{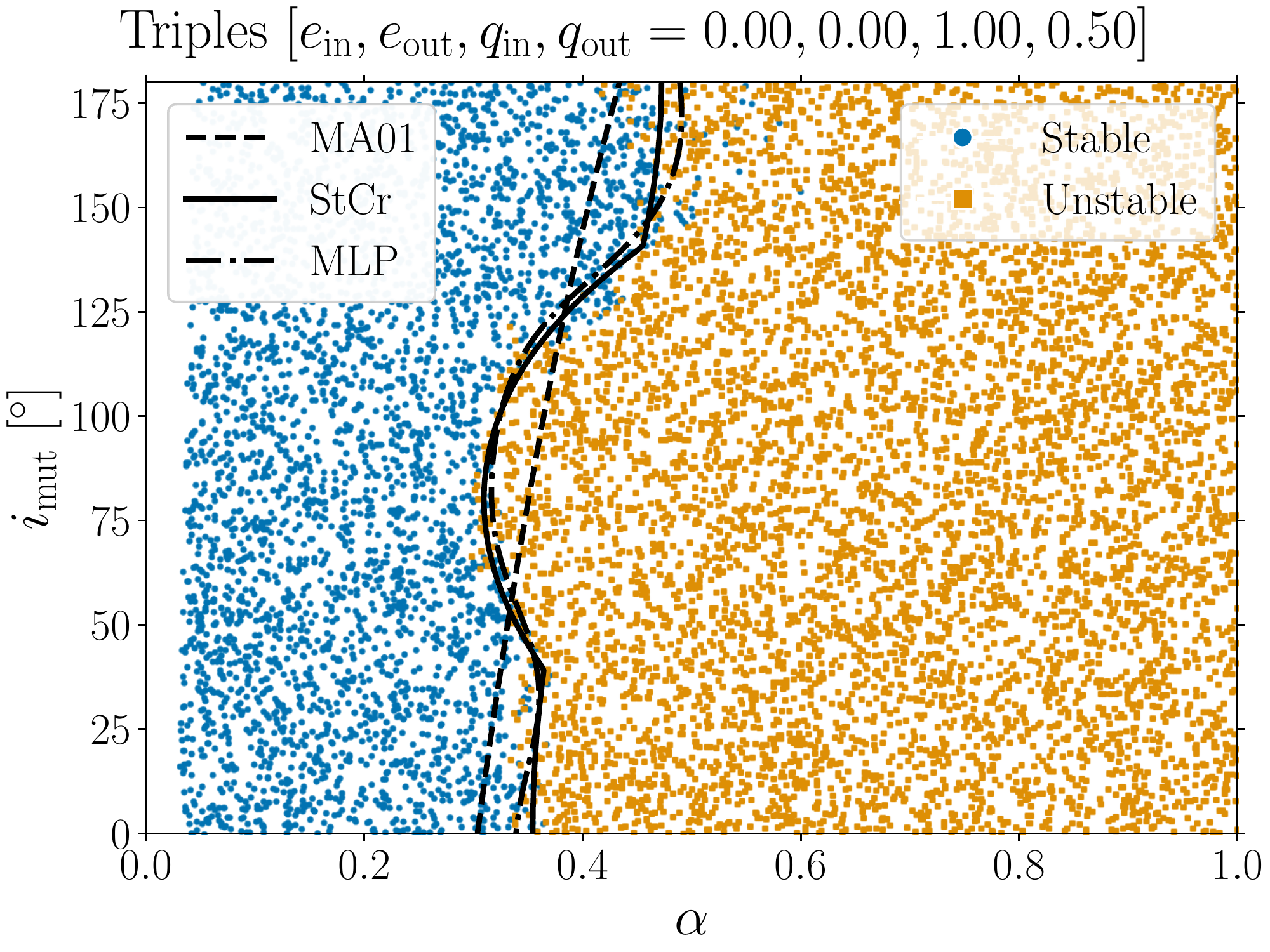}
    \caption{Plot showing a parameter space slice of varying $\alpha$ and $i_{\mathrm{mut}}$. The constant values of the other parameters are mentioned at the top. The left and right panels show the systems which remain stable and become unstable respectively from direct $N$-body simulations. The lines represent the classification boundaries, with the legend labels `MA01', `StCr' and `MLP' referring to the MA01 criterion, our updated stability criterion (Equation~\ref{eqn:new_form}) and our MLP model respectively.}
    \label{fig:ai-1}
\end{figure}

\begin{figure}
	\includegraphics[width=1.0\columnwidth]{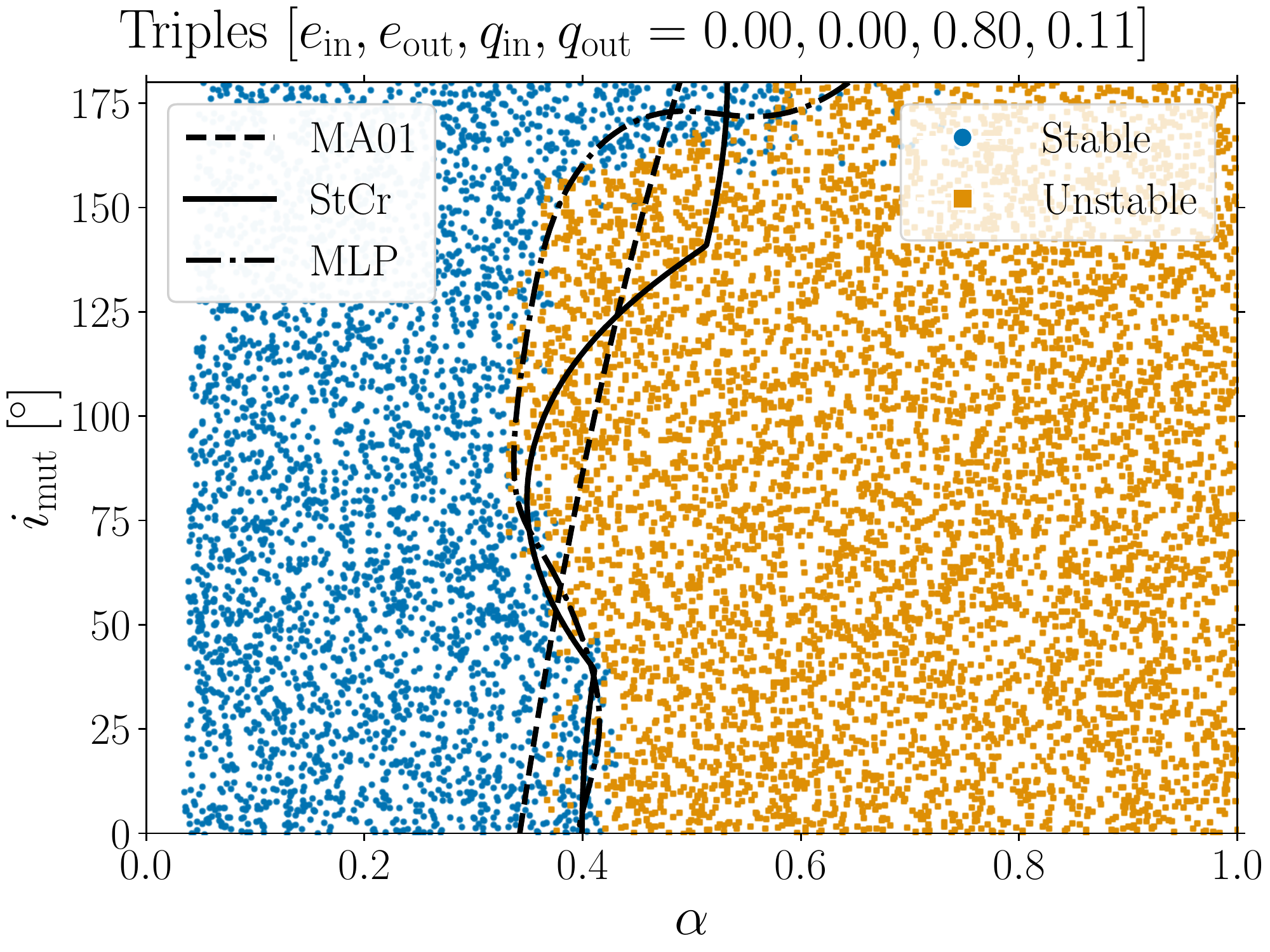}
    \caption{Plot similar to Figure~\ref{fig:ai-1}, but shows a different parameter space slice. The changed constant values are mentioned at the top.}
    \label{fig:ai-8}
\end{figure}

Figures \ref{fig:ai-1} and \ref{fig:ai-8} represent the `bowl-shaped' depressions of instability. Furthermore, the value of $q_{\mathrm{out}}$ also changes the inclination dependence. Equation~\ref{eqn:new_form} works best when $q_{\mathrm{out}}$ is high, i.e., the outer star's mass is comparable or higher than the inner binary stars' masses (Figure~\ref{fig:ai-1}). On the other hand, when $q_{\mathrm{out}}$ is low, mid-range retrograde orbits may be wrongly classified near the stability boundary (Figure~\ref{fig:ai-8}).

\subsection{Mass ratio dependence}
Note that, just like the MA01 formula, Equation~\ref{eqn:new_form} has the same $q_{\mathrm{out}}$ dependence factor $(1 + q_{\mathrm{out}})^{2/5}$ and no dependence on $q_{\mathrm{in}}$. Our parameter space slices of mass ratios showed almost no dependence on $q_{\mathrm{in}}$ (except when mass ratios are very small, $\lesssim$ $0.1$). If we were to replace the quadrupole-order approximated expression for $e_{\mathrm{in,max}}$ by the non-test particle limit formula presented by \citet{2021MNRAS.500.3481H}, the slight dependence on $q_{\mathrm{in}}$ can be accounted for to some extent. However, we found that this replacement did not perform significantly better in classification whereas the expressions are significantly more complicated.

\begin{figure*}
	\includegraphics[width=2.0\columnwidth]{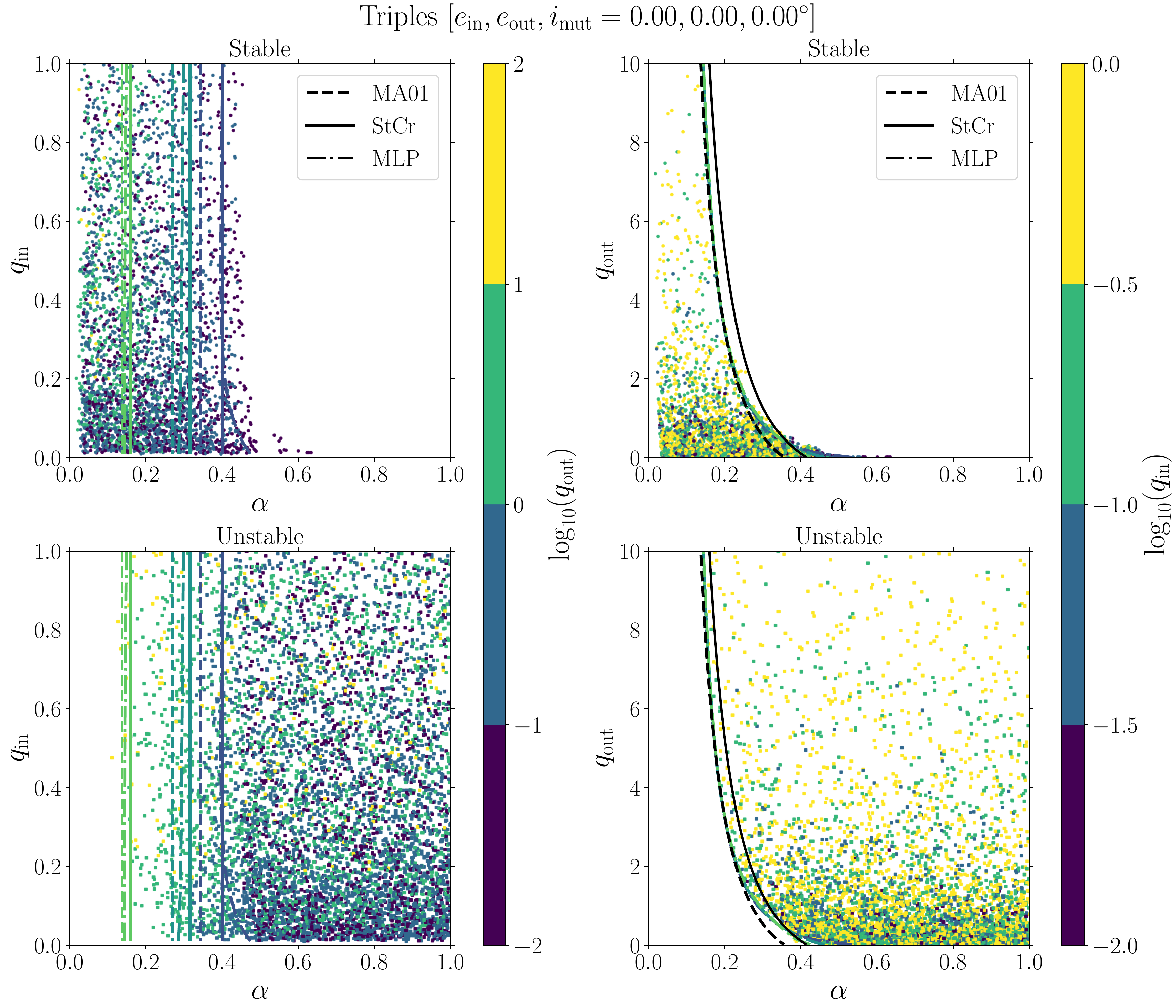}
    \caption{Plot showing a parameter space slice of varying $\alpha$, $q_{\mathrm{in}}$ and $q_{\mathrm{out}}$. The constant values of the other parameters are mentioned at the top. The left panel plots $q_{\mathrm{in}}$ vs $\alpha$, while the right panel plots $q_{\mathrm{out}}$ vs $\alpha$. The missing varied parameter is the colour axis in all panels. The top and bottom panels show the systems which remain stable and become unstable respectively from direct $N$-body simulations. The lines represent contours of the classification boundaries, with the legend labels `MA01', `StCr' and `MLP' referring to the MA01 criterion, our updated stability criterion (Equation~\ref{eqn:new_form}) and our MLP model respectively. The colours of the contour lines coincide with the transition values on the colour axes. The black lines are independent of the colour axis.}
    \label{fig:am-1}
\end{figure*}

Figure \ref{fig:am-1} validates the above claims. The small dependence on $q_{\mathrm{in}}$ can be seen in both figures as a tail in the very-low mass ratio regime.

\section{Machine learning approach} \label{sec:mlp}
Machine learning (ML) classifiers are ubiquitous in the current era of computing. The basic premise of all ML algorithms is to learn from data to improve in classification or regression tasks. The results of such algorithms depend on the model parameters and hyper-parameters, which need to be tuned for reasonable performance.

\subsection{Multi-layer perceptrons}
An artificial neural network (ANN; \citealp{mcculloch1943logical}; for a review, see \citealt{hastie_09_elements-of.statistical-learning}) is a supervised ML algorithm, which uses connected units (neurons), arranged in a series of layers, to perform classification tasks. Each neuron receives inputs from other neurons from the previous layer and provides outputs to the neurons in the following layer. The connection strengths between neurons are controlled by weights. The output of a given neuron is the linear weighted sum of the inputs from the previous layer, which is then passed through an activation function $\phi$ before being input to the next layer. Thus, the output of the $j$th neuron is given as:

\begin{equation}
    y_{j} = \phi \left( \sum_{i=0}^{n} w_{ij} x_{i} + b_{j} \right)
\end{equation}

Here, $\mathbf{x}$ are the inputs from the neurons from the previous layer and $n$ is the number of such neurons. The parameters $\mathbf{w}$ and $\mathbf{b}$ are optimised during training. The network does this by minimising a pre-defined loss function, which depends on the consensus between the true classification labels and the predicted labels.

\begin{figure*}
	\includegraphics[width=1.5\columnwidth]{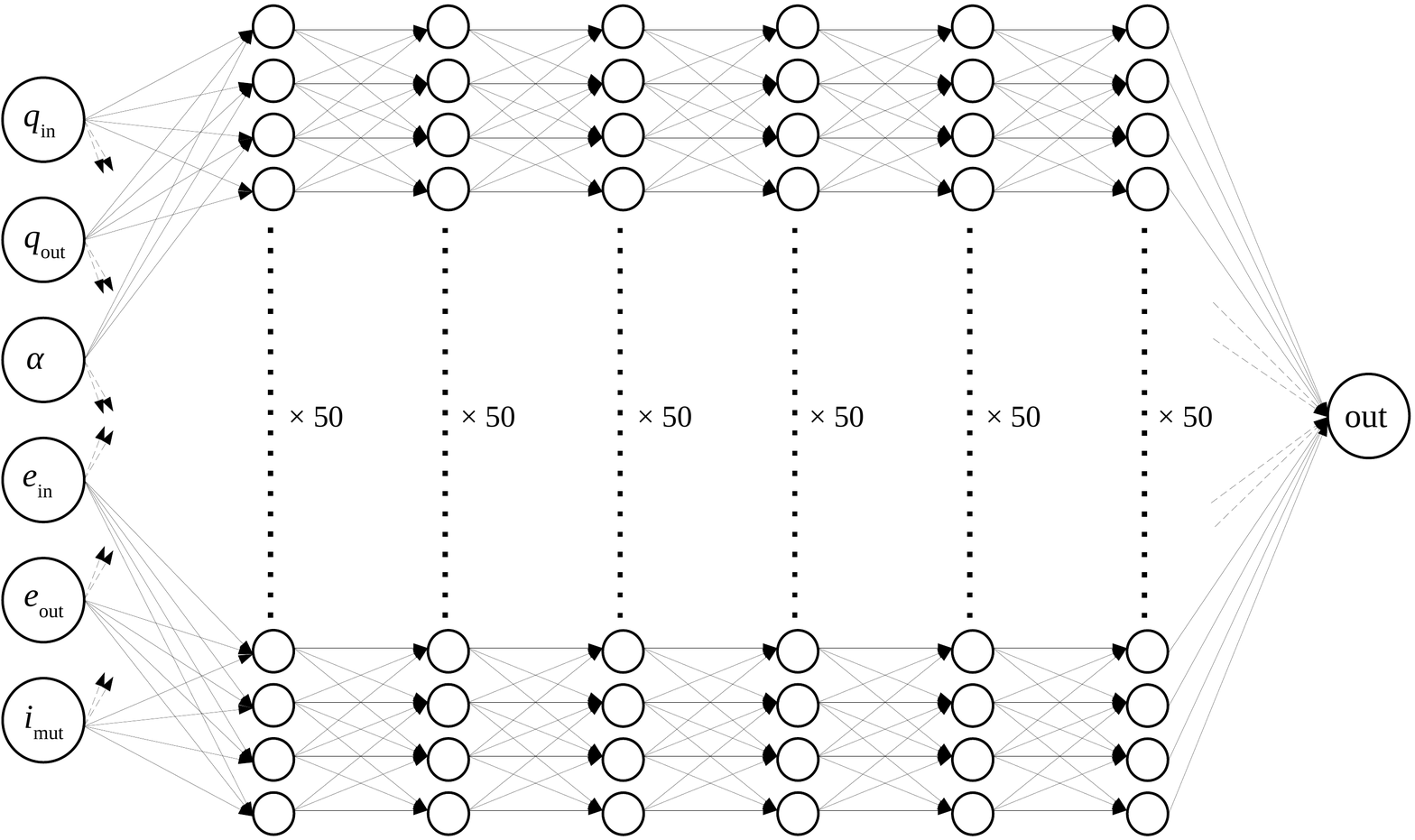}
    \caption{Schematic of the MLP architecture used for classification. Every neuron in a given layer is connected to every neuron in the adjacent layers. The input layer has 6 neurons ($q_{\mathrm{in}}$, $q_{\mathrm{out}}$, $\alpha$, $e_{\mathrm{in}}$, $e_{\mathrm{out}}$, $i_{\mathrm{mut}}$), the output layer has one neuron (value ranges from 0 for `stable' to 1 for `unstable'), and the six hidden layers have 50 neurons each.}
    \label{fig:neuralnet}
\end{figure*}

We specifically implement the simplest form of an ANN called a multi-layer perceptron (MLP; \citealp{Rosenblatt58theperceptron}. MLPs are fully-connected ANNs, in the sense that every neuron in a given layer is connected to every neuron in the immediately adjacent layers. They consist of an input layer, one or more hidden layers and an output layer. Figure~\ref{fig:neuralnet} shows a schematic of the MLP network we used for the classification of stable and unstable triple-star systems.

For the MLP classifier. we use an implementation provided by the \texttt{scikit-learn} \citep{scikit-learn} package in Python.

\subsection{Training and testing data sets}
We generated $10^6$ hierarchical triple-star systems whose initial conditions were sampled uniformly throughout our parameter space, with the parameter limits mentioned in Section~\ref{sec:data}. Besides these six parameters, we also sample the arguments of periapsis and mean anomalies of both orbits uniformly. However, these values are not supplied to the MLP network as inputs. Since the data sparsely populated the six-dimensional space, we added a constraint to ensure that systems are sampled closer to the classification boundary. Additionally, we disregarded systems whose $N$-body integration took longer than five hours. These made up $0.4 \%$ of our total sample.

To do this, we consider the MA01 formula $R_{\mathrm{p,\mathrm{ratio}}} = R_{\mathrm{p}} / R_{\mathrm{p,crit}}$ (Equation~\ref{eqn:MA_form}). If this ratio is greater than 1, a triple-star system is labelled `stable', and vice-versa. However, since the formula does not have 100\% accuracy, there is a range of values around 1 where systems can be classified incorrectly. We found that the range ${0.3 \le R_{\mathrm{p,\mathrm{ratio}}} \le 2.0}$ is the most ambiguous for classification. Thus, after sampling our triple systems, we eliminated any system whose $R_{\mathrm{p,\mathrm{ratio}}}$ value lies outside this range. This ensures that our training data, which lay within the range, is well suited for classification.

During classification, we use an 80:20 split of training and testing data. This, along with 5-fold cross-validation, was implemented to help validate the performance of the classifier.

\subsection{Network architecture and hyper-parameters}
We tried different network architectures and varied the hyper-parameters to arrive at the best-performing MLP model. The performance of a classifier refers to how well the model predicts the labels of previously unseen data -- the test set. We chose our hyper-parameters by training a grid of MLP models, along with 5-fold cross-validation. 

Our final model consists of 6 hidden layers of 50 neurons each (shown in Figure~\ref{fig:neuralnet}). Training a network of this size, with $10^6$ training data points, takes about 30 minutes on 64 cores of an AMD EPYC 7742 CPU. The performance was found to be slightly worse with 4 or 5 hidden layers, and having 10 or 100 neurons in each layer worsened the classification substantially compared to having 50 neurons.

For optimisation, the network uses the \textit{Adam solver} \citep{adamsolver}, which is a stochastic gradient descent algorithm. Finally, we use the \textit{logistic} activation function $\phi(x) = 1  / (1+e^{-x})$ for the neuron outputs. Although slower in implementation than the widely used \textit{ReLu} (rectified linear unit) activation function $\phi(x) = \max(0,x)$, the logistic function performs slightly better.

Other hyper-parameters include:
\begin{itemize}
    \item Batch size, which is the size of the mini-batches used for stochastic gradient descent. After experimenting with batch sizes of 200, 1000 and 5000, we chose 1000.
    \item L2 regularisation term, which is an additional term in the loss function that penalises the squares of the weights. If this penalty is too low, there is an increased risk of over-fitting, and vice-versa. After experimenting with values of $10^{-2}$, $10^{-3}$ and $10^{-4}$, we chose $10^{-3}$.
    \item Initial learning rate, which controls the step-size of updating the weights and biases after each iteration. If it is too low, the optimiser takes a very long time to converge to the minimum of the loss function; if it is too high, the minimum can be skipped. After experimenting with values of $10^{-2}$, $10^{-3}$ and $10^{-4}$, we chose $10^{-2}$.
\end{itemize}

\section{Results and comparison} \label{sec:result}
To quantify the effectiveness of any classification algorithm, we need to consider a few quantities. Some of these include:
\begin{itemize}
    \item Overall score: The fraction of systems which are classified correctly i.e., the `stable' and `unstable' systems are indeed stable and unstable respectively from direct $N$-body simulations.
    \item False stable ($FS$) systems: The number of truly unstable (from $N$-body simulations) systems that the model predicts as `stable'. Similarly, we can define true stable ($TS$) systems.
    \item False unstable ($FU$) systems:  The number of truly stable (from $N$-body simulations) systems that the model predicts as `unstable'. Similarly, we can define true unstable ($TU$) systems.
    \item Precision: The ratio of correctly classified stable (unstable) systems to the total number of systems classified as `stable' (`unstable'): $P_{\mathrm{stable}} = TS/(TS+FS)$ and $P_{\mathrm{unstable}} = TU/(TU+FU)$. It quantifies validity of the classification.
    \item Recall: The ratio of correctly classified stable (unstable) systems to the total number of truly stable (unstable) systems: $R_{\mathrm{stable}} = TS/(TS+FU)$ and $R_{\mathrm{unstable}} = TU/(TU+FS)$. It quantifies completeness of the classification.
\end{itemize}

Table~\ref{tab:score} shows the comparison of scores, precisions and recalls of EK95, MA01, our updated Equation~\ref{eqn:new_form} and our MLP model from Section~\ref{sec:mlp}. It is evident that both our methods score better than the previous studies, and that the MLP model performs the best in classification. It is important to note that these scores can vary depending on the range of sampled parameter space. In our case, we sampled `close' to the MA01 classification boundary line as described in Section~\ref{sec:mlp}.

\begin{table}
	\centering
	\begin{tabular}{llllll}
        \hline
        Classifier & Score & $P_{\mathrm{stable}}$ & $P_{\mathrm{unstable}}$ & $R_{\mathrm{stable}}$ & $R_{\mathrm{unstable}}$ \\
        \hline
        EK95 & 0.86 & 0.96 & 0.81 & 0.68 & 0.98 \\
        MA01 & 0.90 & 0.92 & 0.90 & 0.81 & 0.96 \\
        StCr & 0.93 & 0.93 & 0.92 & 0.85 & 0.96 \\
        MLP & 0.95 & 0.93 & 0.96 & 0.92 & 0.96 \\
        \hline
    \end{tabular}
	\caption{Overall scores, precisions $P$ and recalls $R$ for the different classifiers we compare with. The labels `StCr' and `MLP' refer to our updated stability criterion (Equation~\ref{eqn:new_form}) and our MLP model respectively.}
	\label{tab:score}
\end{table}

\begin{figure}
	\includegraphics[width=1.0\columnwidth]{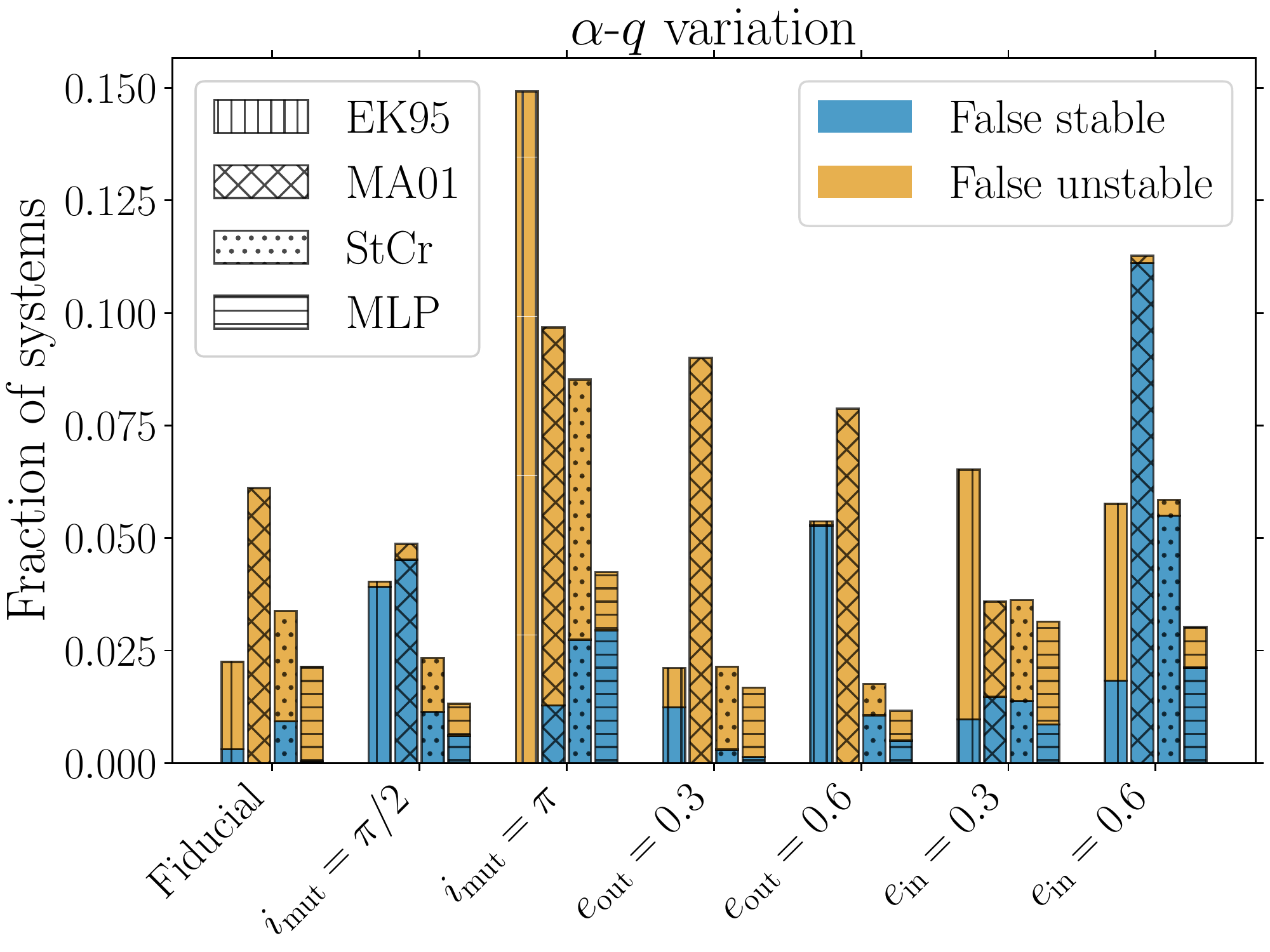}
    \caption{Bar chart illustrating the total fraction of `false stable' and `false unstable' systems (defined in text) in the parameter space slice of varying $\alpha$, $q_{\mathrm{in}}$ and $q_{\mathrm{out}}$. The \textit{x}-axis labels refer to the constant values described in Table~\ref{tab:a-m}. The legend labels `StCr' and `MLP' refer to our updated stability criterion (Equation~\ref{eqn:new_form}) and our MLP model respectively.}
    \label{fig:am_f+-}
\end{figure}

\begin{figure}
	\includegraphics[width=1.0\columnwidth]{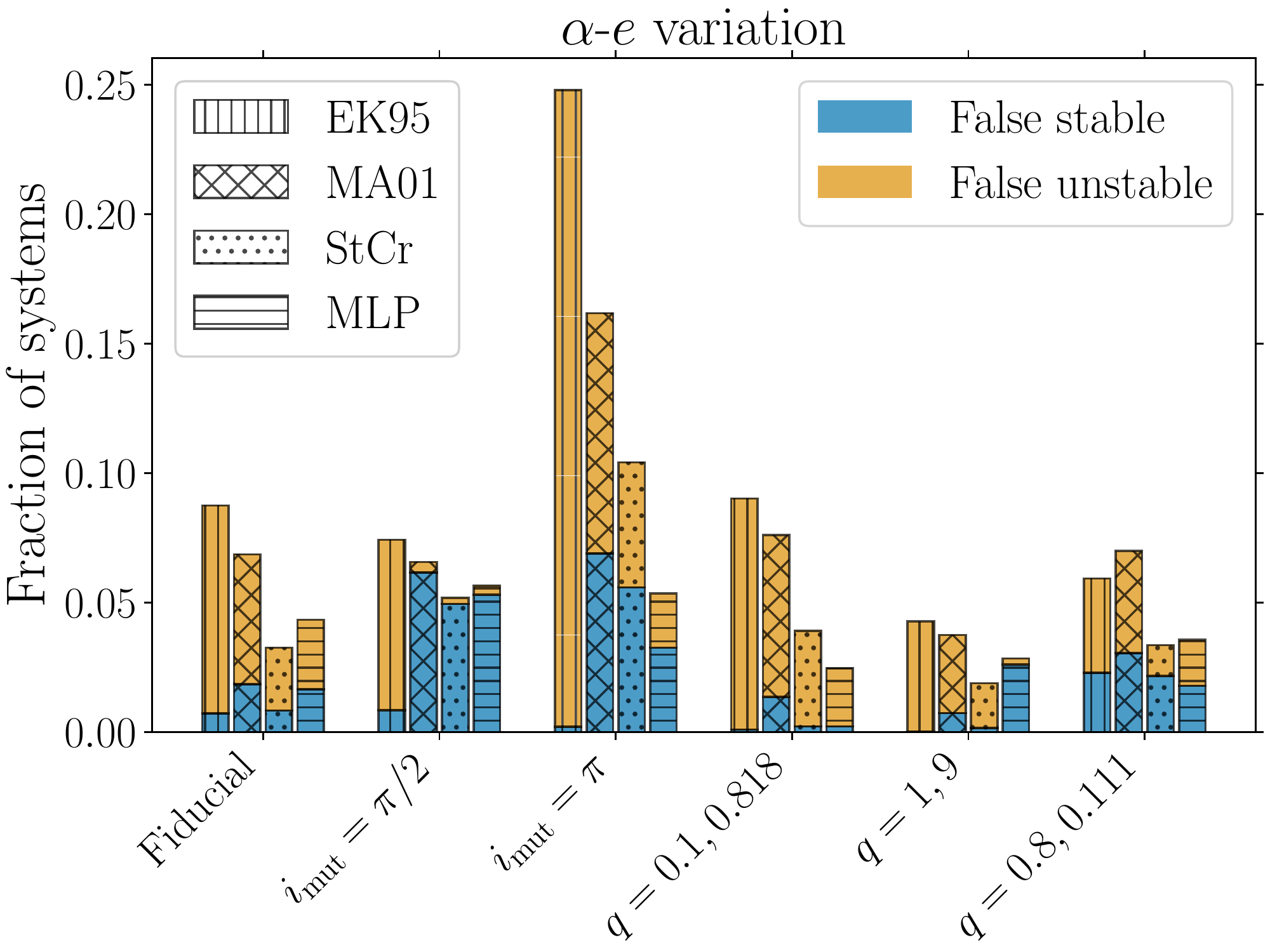}
    \caption{Plot similar to Figure~\ref{fig:am_f+-}, but in the parameter space slice of varying $\alpha$, $e_{\mathrm{in}}$ and $e_{\mathrm{out}}$.}
    \label{fig:ae_f+-}
\end{figure}

\begin{figure}
	\includegraphics[width=1.0\columnwidth]{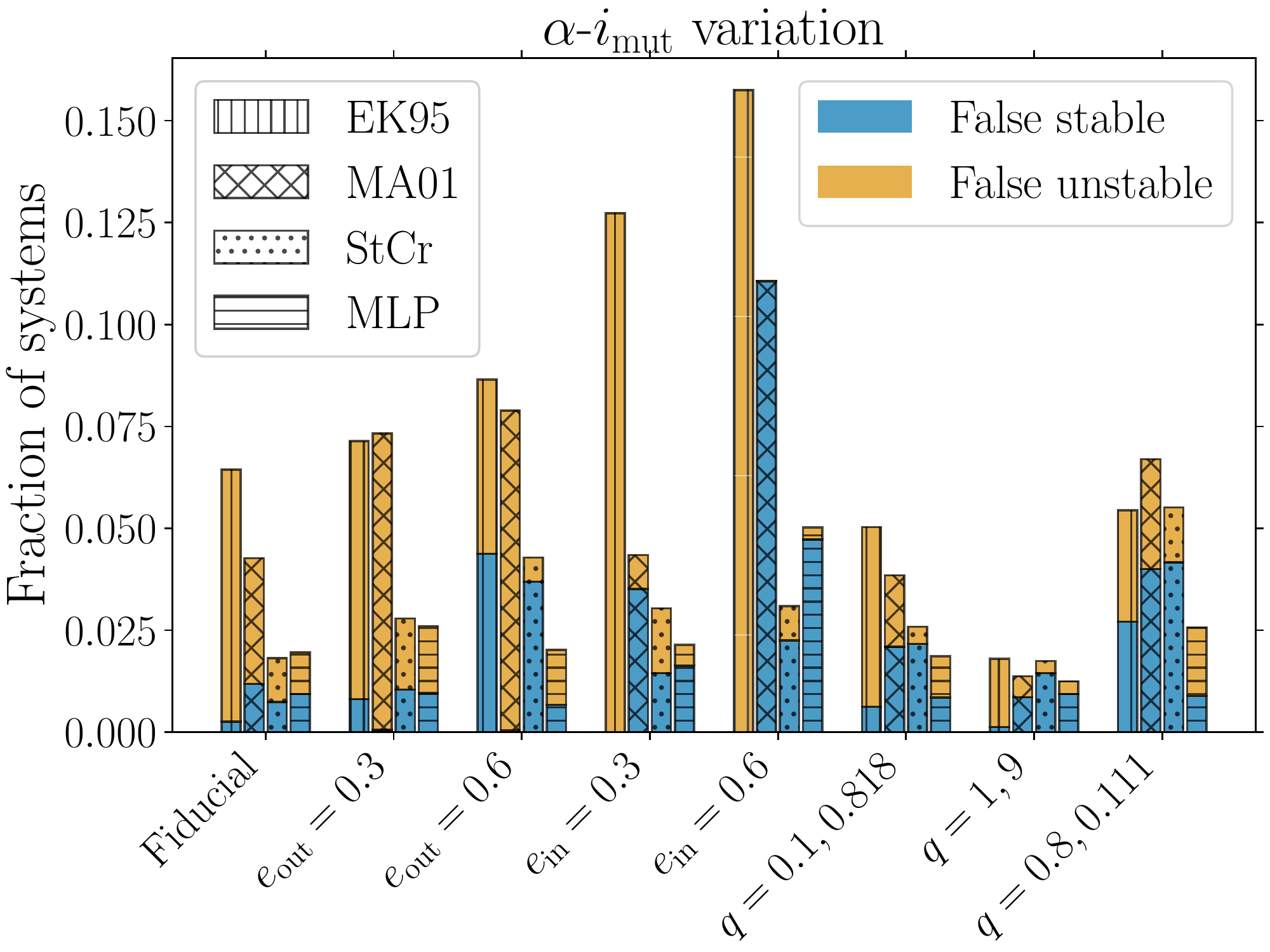}
    \caption{Plot similar to Figure~\ref{fig:am_f+-}, but in the parameter space slice of varying $\alpha$ and $i_{\mathrm{mut}}$.}
    \label{fig:ai_f+-}
\end{figure}

We also show the false stable and unstable rates for the three parameter space slices (see Section~\ref{sec:data}) in Figures \ref{fig:am_f+-}, \ref{fig:ae_f+-} and \ref{fig:ai_f+-} respectively. Again, both our methods perform better than EK95 and MA01 in all parameter space slices, with the MLP model being better in most cases. The figures indicate that the parameter space slices of retrograde orbits ($i_{\mathrm{mut}} = \pi$) have the highest misclassification rates. High $e_{\mathrm{in}}$ values can also lead to less efficient classification. On the other hand, parameter space slices with high $q_\mathrm{out}$ seem to have the best classification rates.
 
\section{Discussion} \label{sec:discuss}

We reiterate that our defining criterion for stability (see Section~\ref{sec:nbody}) is not absolute and that the accuracy of our classification would vary if we were to employ a different defining criterion. Moreover, our integration time of 100 outer orbits does not fully capture the instability of those systems that become unbound much later on. For example, \cite{2008LNP...760...59M} demonstrated that the trajectory of a triple close to the stability boundary can diverge from another almost identical triple with slightly different initial conditions even after 100 outer orbits (see Figure 3.7). Our additional criterion that the inner and outer semimajor axes must not change more than 10\% accounts for some of these unstable systems, but not all. This assumption is reasonable since hierarchical triples that become unstable undergo non-periodic, i.e., persistent transfer of energy between the inner and outer orbits. Since the semimajor axes are inversely proportional to the energies of the respective orbits, any significant change in their values would indicate that the triple is on the verge of becoming unbound.

It is worth noting that, very recently, \cite{2022arXiv220612402L} also presented a machine learning approach to determine the stability of hierarchical triples. They employ a convolutional neural network (CNN), with a time series of orbital parameters as the training data, to perform the classification. This is unlike our classification which depends only on the \textit{initial} values of the parameters, which makes our MLP model simpler and faster to use. They also assume equal masses, whereas we consider the dependence on mass ratios in our analysis. Finally, we highlight that an algebraic criterion like Equation~\ref{eqn:new_form} can be more useful than an ML classifier in understanding the physical dependencies on the initial parameters.

Stability criteria have also been used to classify hierarchical quadruple-star systems. Many studies, including \cite{2021MNRAS.506.5345H} and \cite{2022ApJ...926..195V}, applied the MA01 criterion to quadruples by considering them as two `approximate' triple systems. For example, in a 2+2 quadruple system, one of the two inner binaries can be approximated as a single-star companion to the other inner binary. Hence, applying the triple stability criterion twice (one with each inner binary) yields an effective quadruple-star stability criterion. A similar procedure can be applied on a 3+1 quadruple by approximating the innermost binary as a single star and plugging in the triple stability criterion to the two `nested' triples. Nevertheless, as shown by our preliminary findings, this process does not always predict the stability of the quadruple accurately, especially in cases where the mutual inclinations of the orbits are significant. To overcome this, we are working on a quadruple stability criterion as well (Vynatheya et al. \textit{in prep}) to analyse this problem in greater detail.

\section{Conclusion} \label{sec:conclude}
We presented an algebraic criterion to classify hierarchical triple-star systems based on their dynamical stability. This formula is an improved version of the widely used stability criterion of \cite{2001MNRAS.321..398M} (MA01). We also performed the same classification using a fully connected neural network - a multilayer perceptron (MLP). Our labelled training data of $10^6$ triples are generated by direct $N$-body simulations of triple-star systems using the $N$-body code MSTAR \citep{2020MNRAS.492.4131R}. The main summary and results of the paper are presented in the following points:

\begin{itemize}
    \item Our updated formula, Equation~\ref{eqn:new_form}, adds a dependence on inner orbital eccentricity $e_{\mathrm{in}}$ which is absent in the MA01 formula. It also has a more complicated dependence on mutual orbital inclination $i_{\mathrm{mut}}$, an improvement over the inclination dependence of MA01 described by an ad-hoc inclination factor.
    \item Similar to the MA01 criterion, Equation~\ref{eqn:new_form} does not have an explicit dependence on inner mass ratio $q_{\mathrm{in}}$. This is mostly justified, the only exception being the regime where both inner and outer mass ratios are very low ($\lesssim 0.1$).
    \item Equation~\ref{eqn:new_form} performs better than MA01 in all parameter space slices considered and has an overall classification score of $93\%$. The precisions of stable and unstable systems are $93\%$ and $92\%$ respectively, while the recalls are $85\%$ and $96\%$ respectively. In general, the classification is least effective for retrograde orbits and high inner eccentricities.
    \item Our MLP model is a neural network of six hidden layers of 50 neurons each. It performs even better than Equation~\ref{eqn:new_form} in classification and has an overall classification score of $95\%$. The precisions of stable and unstable systems are $93\%$ and $96\%$ respectively, while the recalls are $92\%$ and $96\%$ respectively.
    \item The classification works best in the parameter ranges of mass ratios $10^{-2} \le q_{\mathrm{in}} \le 1$, $10^{-2} \le q_{\mathrm{out}} \le 10^{2}$ and semimajor axis ratio $10^{-4} < \alpha < 1$. This range corresponds to a major fraction of hierarchical triple-star systems.
    \item Our MLP model is publicly available on Github \href{https://github.com/pavanvyn/triple-stability}{\faGithub} in the form of a simple Python script.
\end{itemize}

\section*{Acknowledgements}
A. S. H. thanks the Max Planck Society for support through a Max Planck Research Group. Funding for the Stellar Astrophysics Centre is provided by The Danish National Research Foundation (Grant agreement no.: DNRF106).

\section*{Data Availability}
The data underlying this article will be shared upon reasonable request to the corresponding author.

\bibliographystyle{mnras}
\bibliography{trip_stable} 

\appendix

\section{Using our MLP model} \label{sec:model}
Using our MLP model is simple. We have uploaded a code on GitHub \href{https://github.com/pavanvyn/triple-stability}{\faGithub} to ensure easy access.

The first step is to install the scikit-learn package (if not already available) using the following terminal command:
\begin{verbatim}
pip3 install scikit-learn
\end{verbatim}

After changing to the repository directory, the \verb|python3| module is run on the terminal as follows:
\begin{verbatim}
python3 mlp_classify.py -qi 1.0 -qo 0.5 -al 0.2 
       -ei 0.0 -eo 0.0 -im 0.0
\end{verbatim}
Here, the arguments \verb|qi|, \verb|qo|, \verb|al|, \verb|ei|, \verb|eo| and \verb|im| refer to $q_{\mathrm{in}}$, $q_{\mathrm{out}}$, $\alpha$, $e_{\mathrm{in}}$, $e_{\mathrm{out}}$ and $i_{\mathrm{mut}}$ respectively. The parameter ranges should be restricted to the values given in Section~\ref{sec:data} for optimal results.

It is also possible to import the MLP classifier to another custom \verb|python3| script. The input parameters can also be \verb|numpy| arrays, as shown in the sample script below:
\begin{verbatim}
import numpy as np
from mlp_classify import mlp_classifier

# generate initial numpy arrays qi, qo, al, ei, eo, im

mlp_pfile = "./mlp_model_best.pkl"

mlp_stable = mlp_classifier(mlp_pfile, qi, qo, 
           al, ei, eo, im)

# returns True if stable, False if unstable
\end{verbatim}

\bsp	
\label{lastpage}
\end{document}